\documentclass{article}

\usepackage[final,nonatbib]{nips_2018}

\usepackage[utf8]{inputenc} 
\usepackage[T1]{fontenc}    
\usepackage{hyperref}       
\usepackage{url}            
\usepackage{booktabs}       
\usepackage{amsfonts}       
\usepackage{nicefrac}       
\usepackage{microtype}      
\usepackage{graphicx}
\usepackage{color}
\usepackage{subfigure}

\title{Covering the News with (AI) Style}
\vspace{-0.8cm}
\author{
  Michele Merler \hspace{2cm} Cicero Nogueira dos Santos \hspace{2cm}  Mauro Martino 
  \And Alfio M. Gliozzo \hspace{2cm}  John R. Smith
  \\ \\
  \textbf{IBM Research AI}\\
  New York, NY\\
  \texttt{\{mimerler,cicerons,mmartino,gliozzo,jsmith\}@us.ibm.com} 
}

\begin{document}

\maketitle

\vspace{-0.8cm}

\begin{figure}[h]
    \centering
    \includegraphics[width=0.7\linewidth]{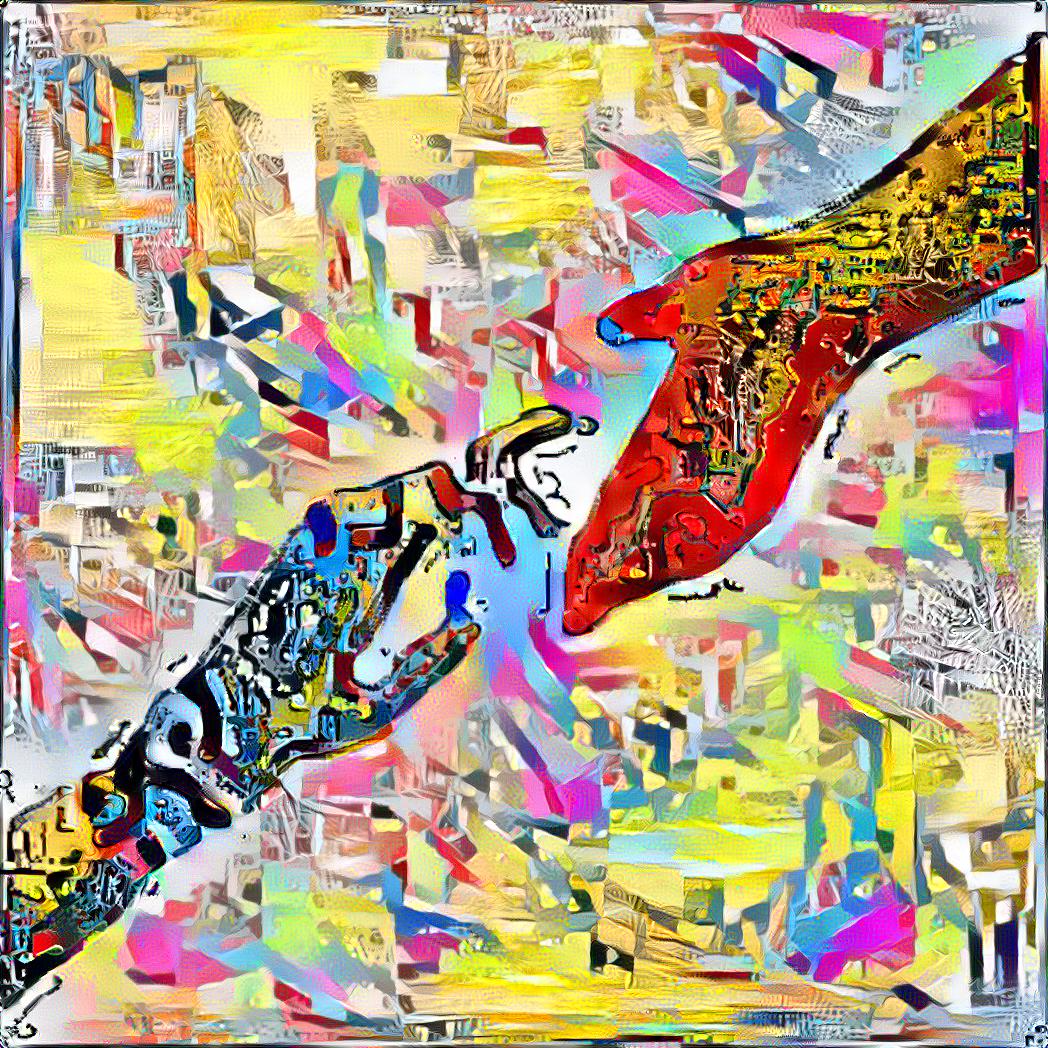}
    \vspace{-0.2cm}
    \caption{AI-generated image of AI  for the New York Times special session on AI}
    \label{fig:cover}
\end{figure}

\begin{figure*}[t]
    \centering
    \includegraphics[width=1\linewidth]{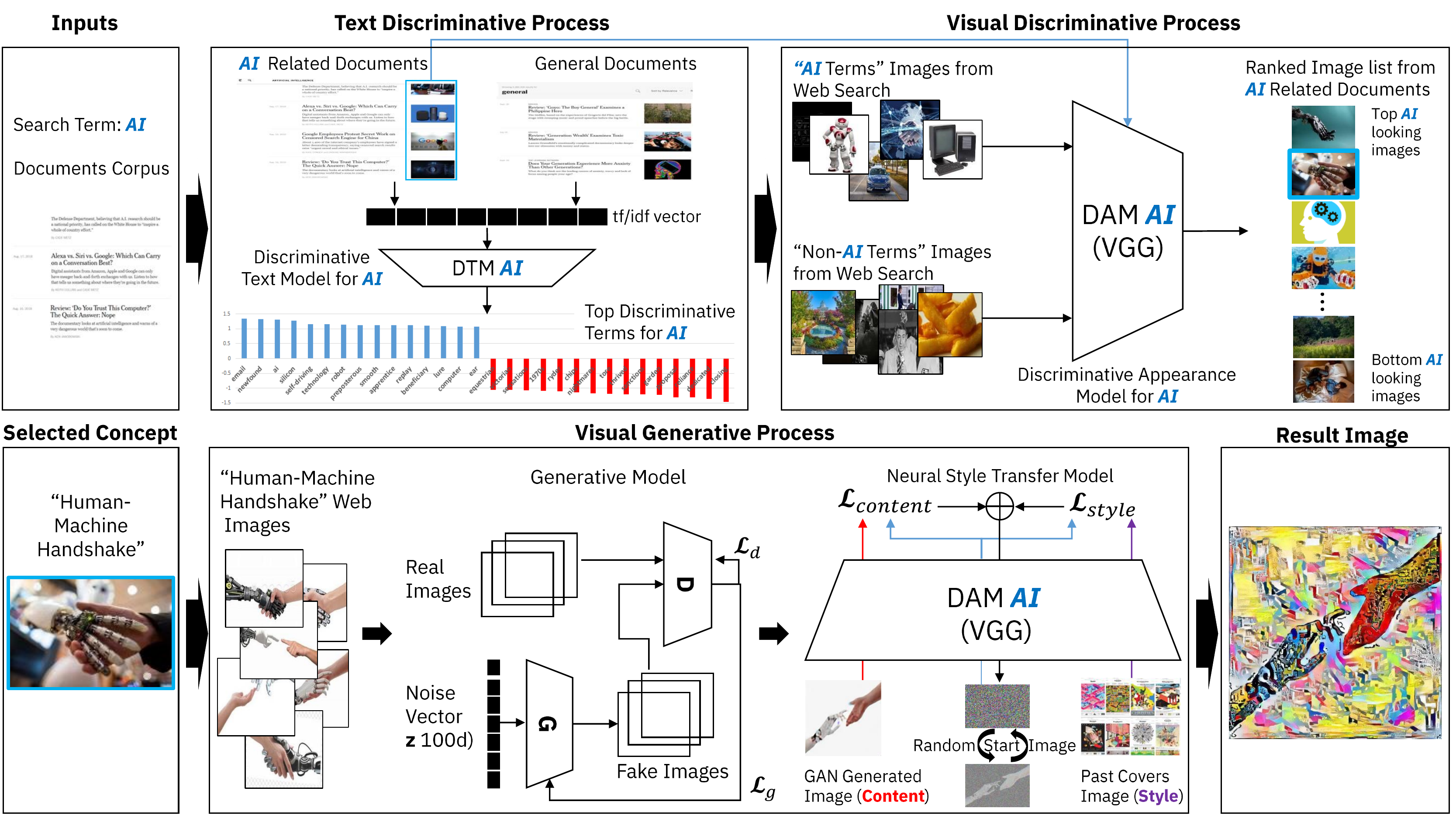}
    \vspace{-0.5cm}
    \caption{Pipeline of the Art Generation Process}
    \label{fig:diagram}
\end{figure*}

We introduce a multi-modal discriminative and generative framework capable of assisting humans in producing visual content related to a given theme, starting from a collection of documents (textual, visual, or both). This framework can be used by editors to generate images for articles, as well as books or music album covers.
Motivated by a request from the The New York Times (NYT) seeking help to use AI to create art for their special section on Artificial Intelligence, we demonstrated the application of our system in producing such image. The result is presented in Figure \ref{fig:cover}\footnote{This image was published in page F7 of the NYT issue of October 19th, 2018} \footnote{``AI Self Portrait'' was selected in the art gallery of the NeurIPS Workshop on Machine Learning for Creativity 2018 (\url{https://bit.ly/32HuHiJ})}.
Starting from a corpus of documents and a keyword term, the goal of our system is to generate an image reflecting those inputs. We accomplish it with the pipeline described in Figure \ref{fig:diagram}, which involves three major steps. In the following we describe the steps in detail, using the NYT ``AI'' image generation as an example use case of the system.

\textbf{Identify a core visual concept}. 
As an input corpus we collected about 3,000 articles from the NYT archives, half related to ``AI'' and half unrelated to it (to do so we use the metadata provided by the publisher). Then, we applied basic natural language processing tools to identify terms for each document and we used this representation to train a discriminative text model (DTM)a. Next, we used the dimensions in the tf/idf vector space with the highest weights according to DTM and used them to identify the top-30 most discriminative semantic concepts, 15 positives and 15negatives. Those terms were used to collect images using a web image search engine, on top of which a VGG16 visual recognition deep network was trained to visually determine in discriminative fashion "AI" appearance (DAM AI model). 
Then, we applied the DAM AI network to rank images from the AI related NYT articles according to their strength of depicting or representing “AI”. Finally, the human was included in the loop to select one of the top ranked images: an image of a human and robot shaking hands. This became the basis concept of the visual generative process.


\begin{figure*}[ht]
    \centering
    \subfigure[GFMN Examples 1]{
    \includegraphics[width=0.49\linewidth]{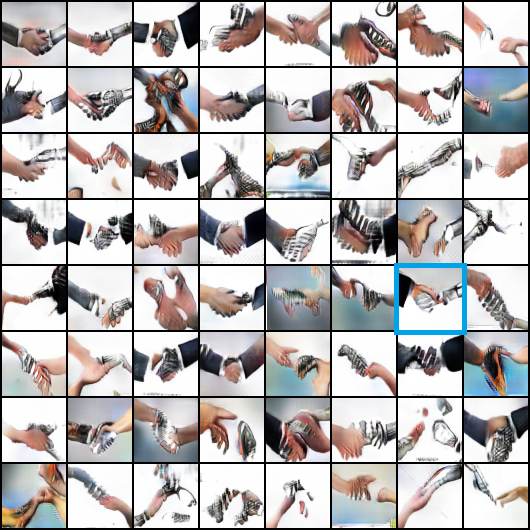}}
     \subfigure[GFMN Examples 2]{
    \includegraphics[width=0.49\linewidth]{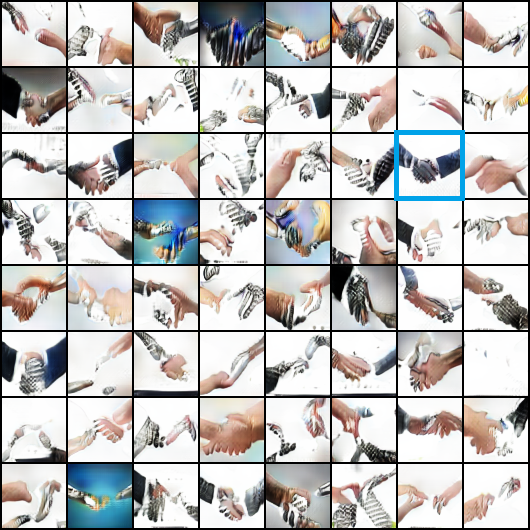}}
    \subfigure[Style Transfer GFMN Example 1]{
    \includegraphics[width=0.49\linewidth]{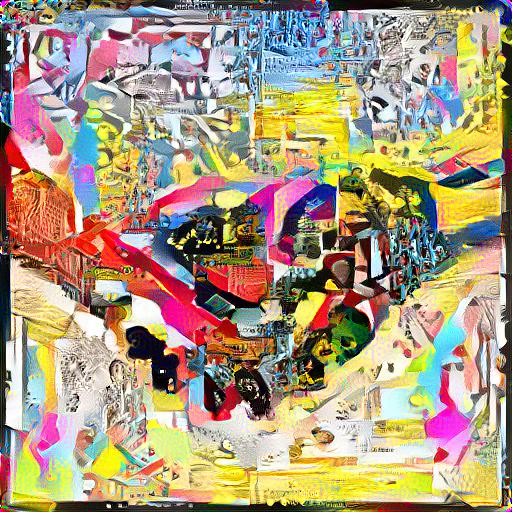}}
     \subfigure[Style Transfer GFMN Example 2]{
    \includegraphics[width=0.49\linewidth]{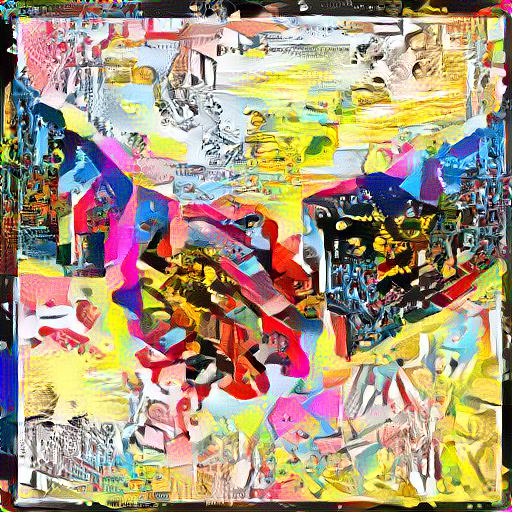}}
   \caption{Examples of Human-robot handshake Images  (resolution 64x64 pixels) generated using GFMN and respective style transfer results.}
   \label{fig:style_transfer4}
\end{figure*}

\textbf{Create an original image that captures the given concept}.
Using "Human-Machine Handshake" as a query concept, we built a training dataset of approximately 1,000 images of human and robot hands collected from the web. We then trained a generative neural network (GAN) to draw new images of human and robot hands. Specifically for our use case we used a BEGAN \cite{began_ARXIV17} model, which ran for 17K iterations and for the generator used a random input vector $\textbf{z}$ of dimensionality 100. Batch size was 16, image dimensions 128x128, learning rate 1e-4, $\gamma=0.5$ and Adam optimizer. Other generative models were tested as well (see Supplementary Material Figures \ref{fig:style_transfer3} and \ref{fig:style_transfer4}).

\textbf{Present the image in a way that fits the magazine visual style}.
 The GAN generated images at this stage present two limitations from a publishing perspective: resolution (128x128, we did not try high resolution GANs \cite{karras2018progressive} at the time of this experiment) and style (the magazine content tend to be more graphic and less photo-realistic (see detail in Supplemental Material Figure \ref{fig:style_transfer}(a))). In order to solve both issues, we adopted a neural style transfer approach \cite{gatys2015neural}. Instead of transferring the style of a single image, we collected a sample of cover art from the NYT, tiled them together in a single style reference image, and trained a style transfer network with it. The base network for the style transfer was the same DAM VGG16. Style was optimized for reconstruction in layers $conv1\_1$, $conv2\_1$, $conv3\_1$, $conv4\_1$ and $conv5\_1$, while content (any of the BEGAN-generated handshake images) was optimized for layer $conv4\_2$.
 In that way, we were able to produce images of "AI" matching the NYT “visual language” for cover art. Furthermore, the output resolution of the style-transferred image could be as high as 1048x1048. Finally, we chose among multiple generations the image shown in Figure \ref{fig:cover} based on overall concept clarity and artistic style.

In this specific work, we have opted to keep the human involved in various stages of the creative process, since we view our system as an assistive editorial tool more than fully self-contained. However, the entire pipeline could be automated. For example, one could use an image captioning system \cite{showandtell_CVPR15} to automatically determine the concept out of the top ranked images by the DAM model. Furthermore, components of the pipeline can be substituted or re-arranged during the creative process. For example, one of the selected images could be directly employed as input to the neural style transfer model (as illustrated in Supplementary Material Figure \ref{fig:style_transfer}), instead of using a fully generative solution.
We envision this type of solution to be helpful for editorial purposes in general, as it could be applied for example to book or music cover generation, perhaps even employing acoustics analysis as well. Inserting tools capable of evoking emotions \cite{AlvarezMelis2017TheEG} as well as style can also be an interesting future direction.




\medskip

\small

\vfill
\pagebreak

\section{Supplementary Material} \label{sec:supplementary}

Here we show some alternative designs and examples for the AI generated images using either style transfer on selected existing images, or different GAN frameworks utilized to generated human-robot handshake images.

\begin{figure*}[ht]
    \centering
    \subfigure[NYT Covers Style Image]{
    \includegraphics[width=0.46\linewidth]{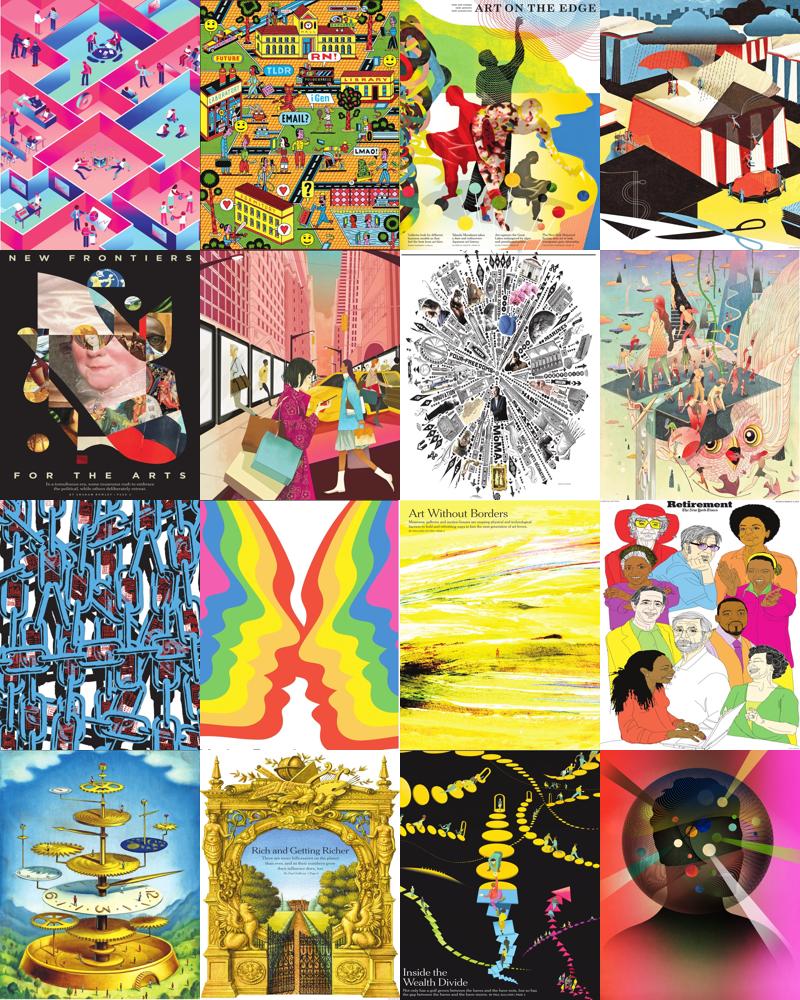}}
     \subfigure[Top Ranked Images from NYT AI-related Articles]{
    \includegraphics[width=0.52\linewidth]{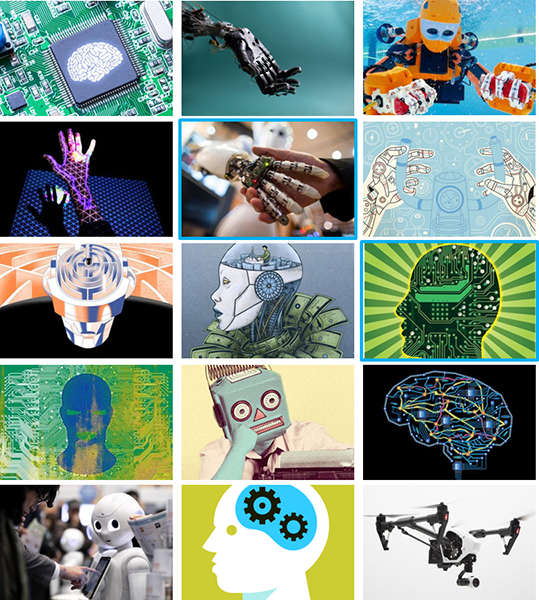}}
    \subfigure[Style Transfer Selection Example 1]{
    \includegraphics[width=0.49\linewidth]{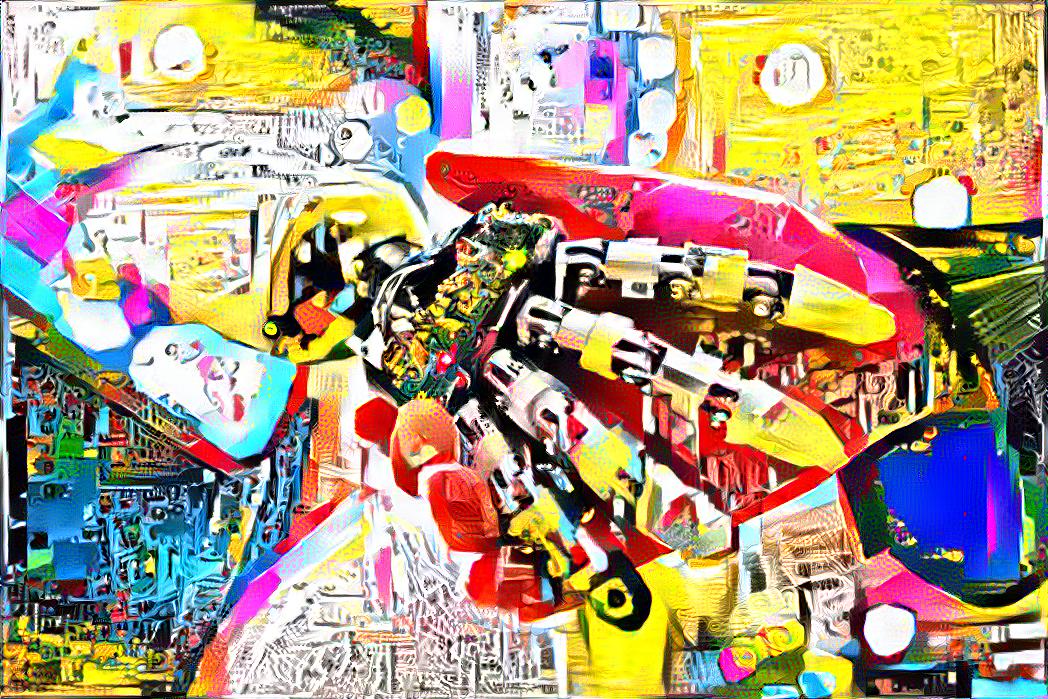}}
     \subfigure[Style Transfer Selection Example 2]{
    \includegraphics[width=0.49\linewidth]{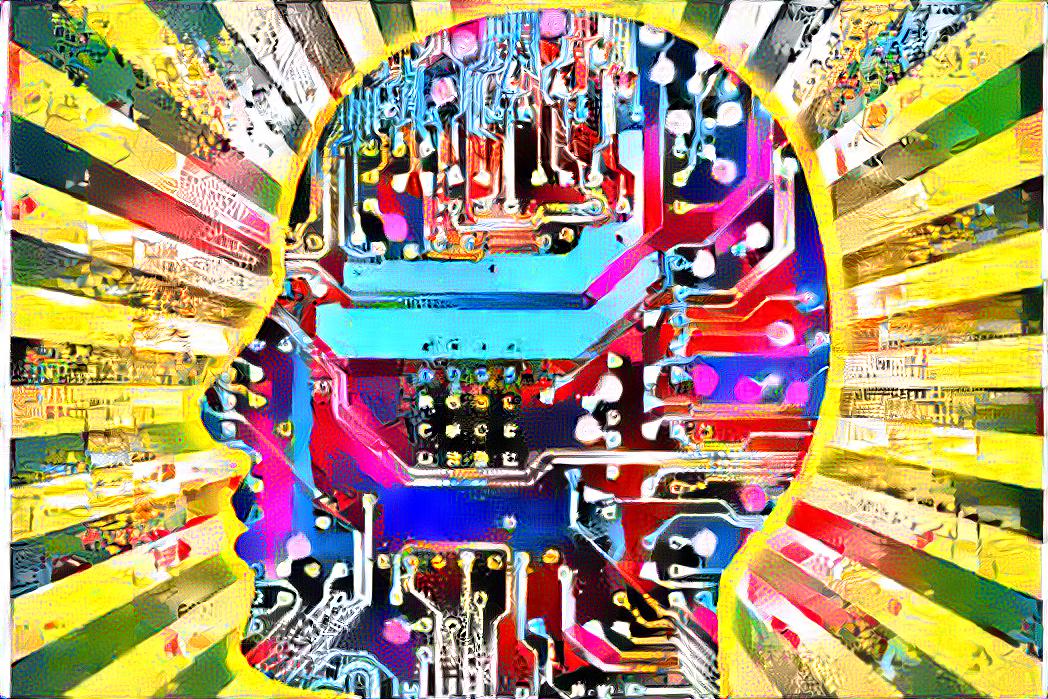}}
   \caption{NYT covers style (a), top ranked AI-looking images from NYT articles according to the DAM AI model (b) and examples of top ranked images style transfer results(c,d).}
   \label{fig:style_transfer}
\end{figure*}

\begin{figure*}[ht]
    \centering
    \subfigure[BEGAN Example 1 (base for Figure\ref{fig:cover})]{
    \includegraphics[width=0.31\linewidth]{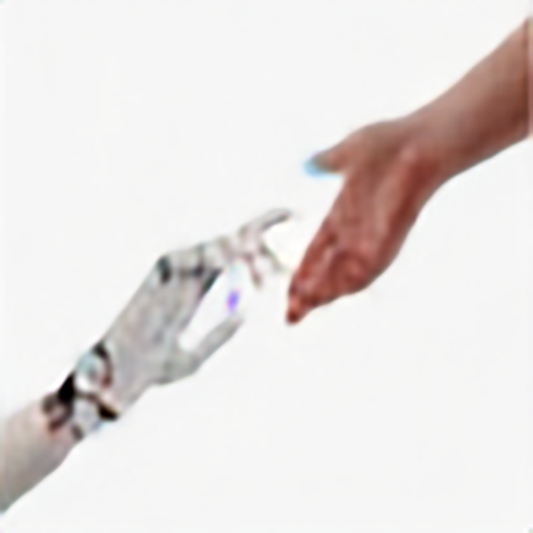}}
     \subfigure[BEGAN Example 2]{
    \includegraphics[width=0.31\linewidth]{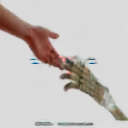}}
    \subfigure[BEGAN Example 3]{
    \includegraphics[width=0.31\linewidth]{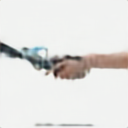}}
    \subfigure[Style Transfer BEGAN Example 2]{
    \includegraphics[width=0.49\linewidth]{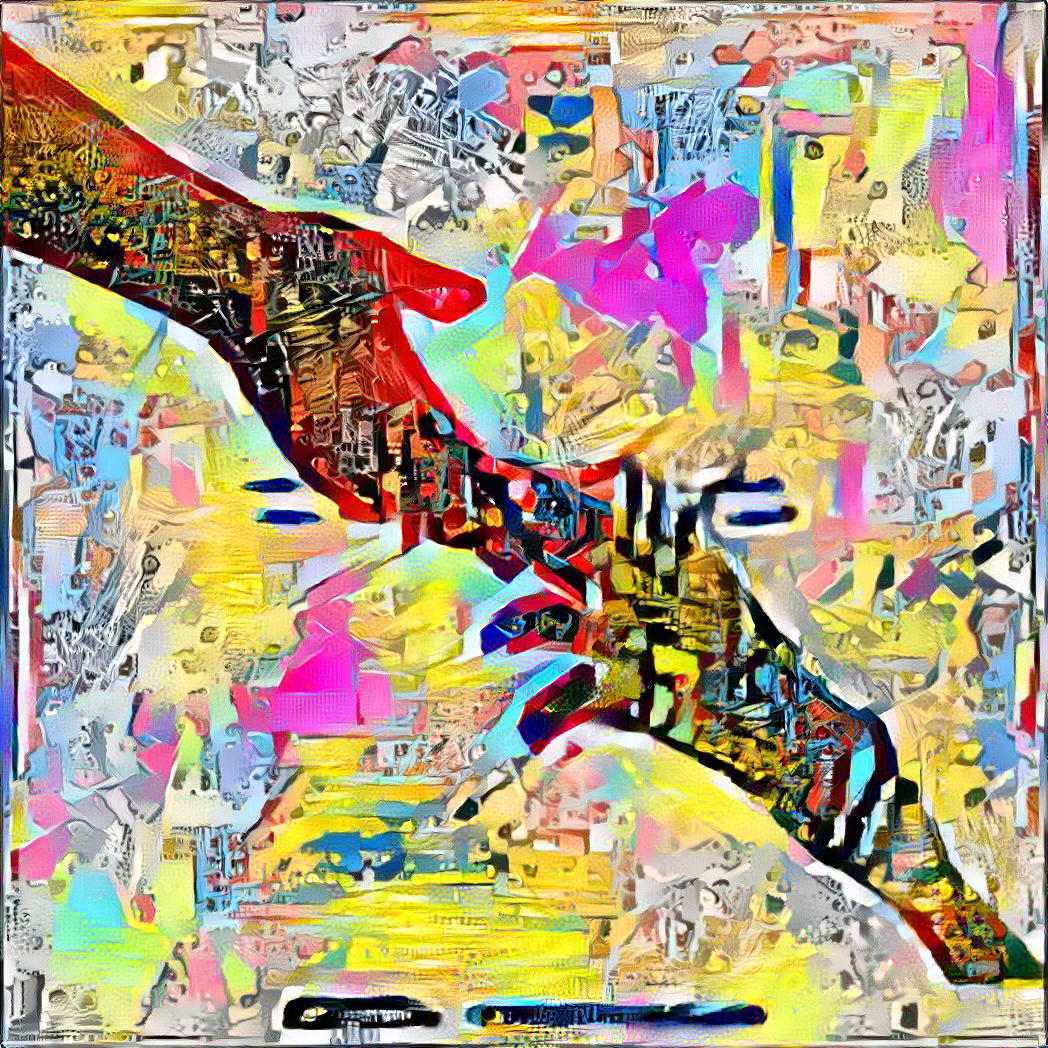}}
     \subfigure[Style Transfer BEGAN Example 3]{
    \includegraphics[width=0.49\linewidth]{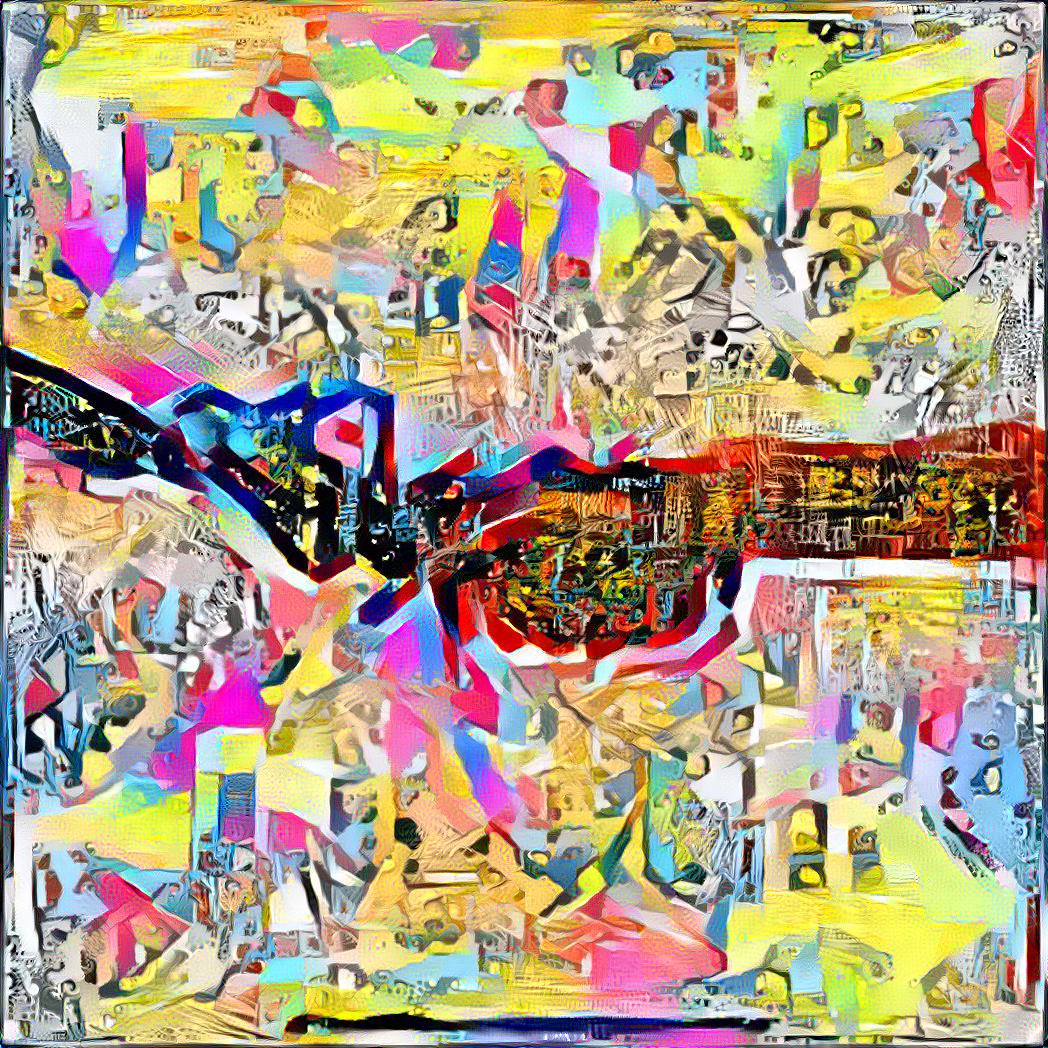}}
   \caption{Examples of BEGAN generated Human-robot handshake Images (resolution 128x128 pixels) and respective style transfer results.}
   \label{fig:style_transfer2}
\end{figure*}

\begin{figure*}[ht]
    \centering
    \subfigure[GFMN Example 1]{
    \includegraphics[width=0.32\linewidth]{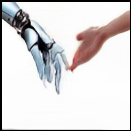}}
     \subfigure[GFMN Example 2]{
    \includegraphics[width=0.32\linewidth]{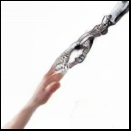}}
    \subfigure[GFMN Example 3]{
    \includegraphics[width=0.32\linewidth]{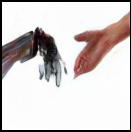}}
    \subfigure[Style Transfer GFMN Example 1]{
    \includegraphics[width=0.49\linewidth]{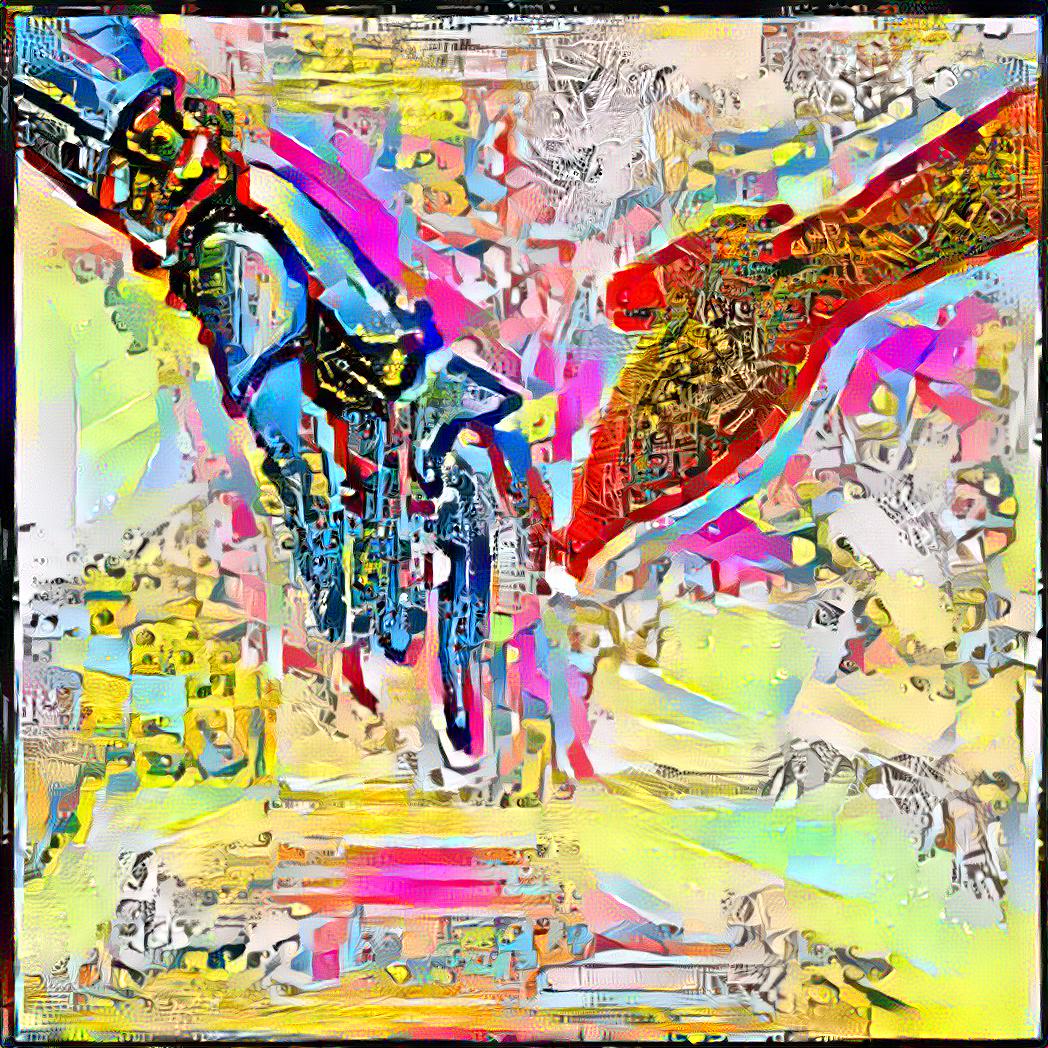}}
     \subfigure[Style Transfer GFMN Example 3]{
    \includegraphics[width=0.49\linewidth]{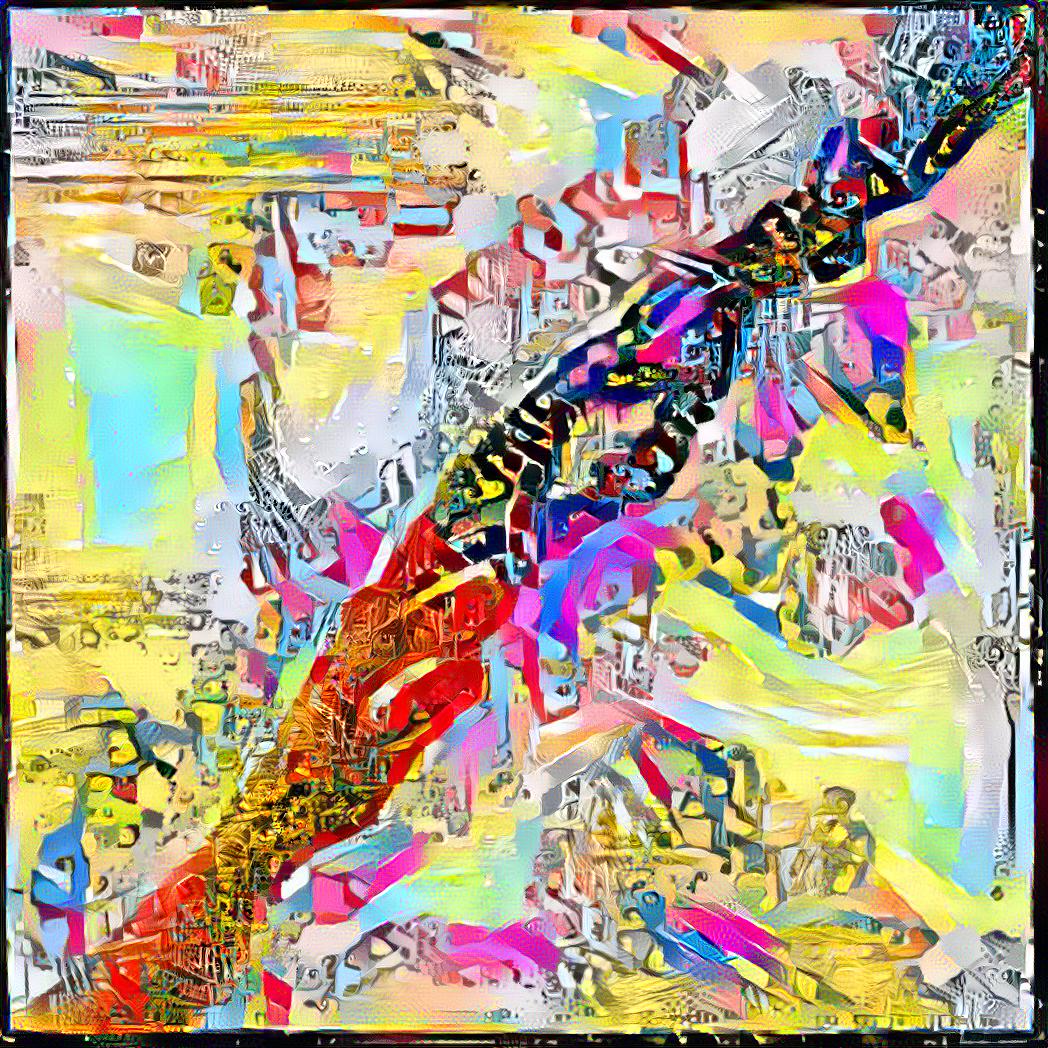}}
   \caption{Examples of Generative Feature Matching Networks (GFMN) \cite{anonymous2019gfmn} generated Human-robot handshake Images (resolution 128x128 pixels) and respective style transfer results.}
   \label{fig:style_transfer3}
\end{figure*}

\begin{figure*}[ht]
    \centering
    \subfigure[GAN Example 1]{
    \includegraphics[width=0.49\linewidth]{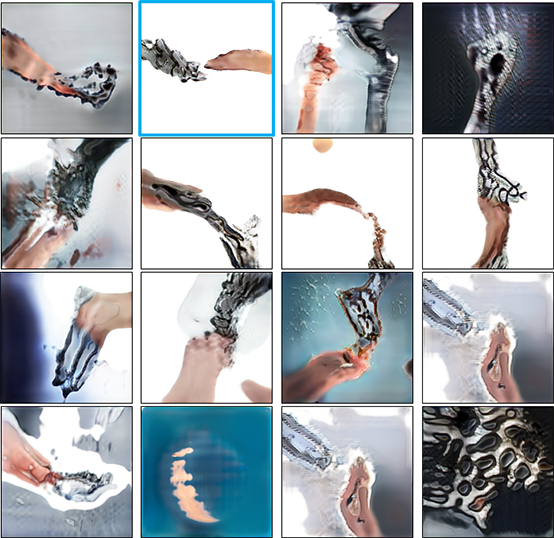}}
     \subfigure[GAN Example 2]{
    \includegraphics[width=0.48\linewidth]{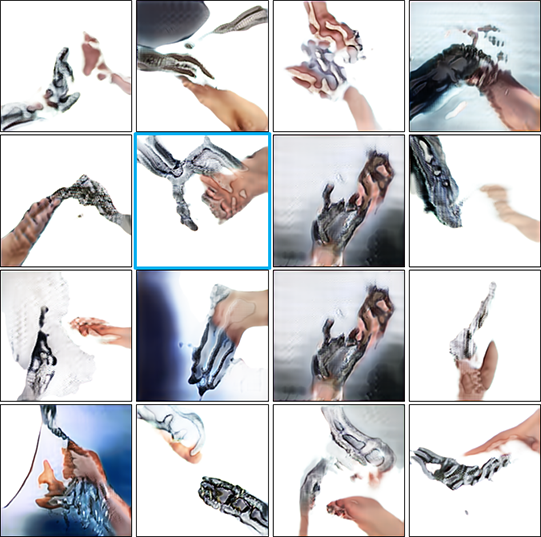}}
    \subfigure[Style Transfer GAN Example 2]{
    \includegraphics[width=0.49\linewidth]{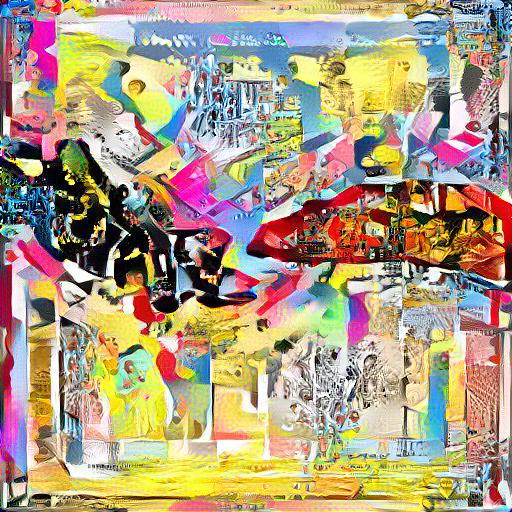}}
     \subfigure[Style Transfer GAN Example 3]{
    \includegraphics[width=0.49\linewidth]{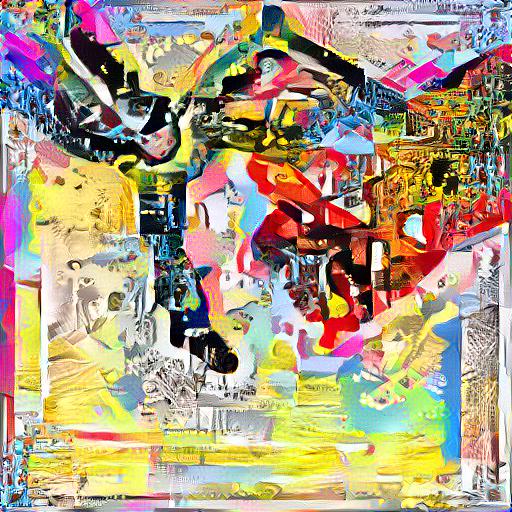}}
   \caption{Examples of DCGAN \cite{dcgan_ICLR16} generated Human-robot handshake Images (resolution 256x256 pixels) and respective style transfer results.}
   \label{fig:style_transfer5}
\end{figure*}

\end{document}